\documentclass[runningheads]{llncs}
\usepackage[T1]{fontenc}
\usepackage{amsfonts}
\usepackage{graphicx}
\usepackage{hyperref}
\usepackage{color}

\begin{document}
\title{Towards Safety-Compliant Transformer Architectures for Automotive Systems}
\author{Sven Kirchner\inst{1}\orcidID{0000-0001-5764-4980} \and 
Nils Purschke\inst{1}\orcidID{0009-0008-9470-5795} \and
Chengdong Wu\inst{1}\orcidID{0009-0002-0445-4934} \and
Alois Knoll\inst{1}\orcidID{0000-0003-4840-076X}}
\authorrunning{S. Kirchner et al.}
\institute{Technical University of Munich, Chair of  Robotics, Artificial Intelligence (AI) and Embedded Systems, Boltzmannstraße 3, 85748 Garching bei München, Germany
\email{sven.kirchner@tum.de}\\
}
\maketitle
%


\begin{abstract}
Transformer-based architectures have shown remarkable performance in vision and language tasks but pose unique challenges for safety-critical applications. This paper presents a conceptual framework for integrating Transformers into automotive systems from a safety perspective. We outline how multimodal Foundation Models can leverage sensor diversity and redundancy to improve fault tolerance and robustness. Our proposed architecture combines multiple independent modality-specific encoders that fuse their representations into a shared latent space, supporting fail-operational behavior if one modality degrades. We demonstrate how different input modalities could be fused in order to maintain consistent scene understanding. By structurally embedding redundancy and diversity at the representational level, this approach bridges the gap between modern deep learning and established functional safety practices, paving the way for certifiable AI systems in autonomous driving.
\end{abstract}


\section{Introduction}

\subsection{Classic Safety in Automotive}
Traditional linear development processes, such as the V-model or the waterfall model~\cite{pressman2009}, have long been used in software engineering but offer limited flexibility for addressing the demands of modern machine learning systems. To manage the increasing complexity of safety-critical applications like autonomous driving, the automotive industry has published various safety standards to ensure functional safety and robustness. ISO26262\cite{ISO26262} defines a comprehensive framework for addressing functional safety in road vehicles. It describes systematic hazard analysis and risk assessment procedures for deriving safety goals and the corresponding Automotive Safety Integrity Levels (ASILs). These levels dictate the necessary rigour of analysis, design and verification activities~\cite{Koopman2019}.

One critical concept in ISO~26262 is ASIL decomposition, which enables the reduction of overall system safety requirements through redundancy and diversification. Prior work has demonstrated that ASIL decomposition can be implemented at runtime by verifying the independence and diversity of redundant subsystems, which deliver signals with reduced ASIL levels ~\cite{Frtunikj2014}. This facilitates the safe integration of new hardware or software components into open automotive systems through dynamic assessment and configuration of appropriate monitoring and voting mechanisms.

\subsection{Transformer-Based Architectures}

In parallel with advances in automotive safety engineering, recent breakthroughs in machine learning have shifted the focus from traditional coding to training large models. The introduction of the Transformer architecture ~\cite{vaswani2017} revolutionized natural language processing by leveraging self-attention and multi-head attention mechanisms to capture long-range dependencies in data. Extending these concepts to the visual domain, the Vision Transformer (ViT)~\cite{dosovitskiy2020} demonstrated that Transformer-based models could match or surpass convolutional neural networks (CNNs) \cite{LeCun1989} in image recognition tasks by treating images as sequences of patches.

Building on this foundation, large vision–language models (VLMs) have emerged as powerful tools for joint understanding of visual and textual information. For example, FLAVA~\cite{Singh2022} uses a single shared Transformer encoder to fuse vision and language modalities, while VisualBERT~\cite{li2019} combines an image encoder with a text encoder based on BERT~\cite{devlin2019} to enable multimodal reasoning.

\subsection{Multimodal VLMs and Redundancy through Sensor Diversity}

Recent research has expanded the capabilities of VLMs by integrating additional modalities, such as depth and LiDAR data, to improve scene comprehension and robustness. Research has shown that monocular metric depth estimation has the potential to transform large-scale 2D internet images into 3D point clouds. This enables a richer understanding of 3D scenes within a VLM framework \cite{Chen2024}. Another approach introduces LiDAR point clouds into large language models by collapsing the 3D data along the z-axis to create bird's-eye-view (BEV) feature maps, which are then processed by view-aware Transformer modules \cite{Yang2025}.


\section{Methodology}

\begin{figure*}[htb]
    \vspace{1em}
    \centering
    \includegraphics[width=0.96\textwidth]{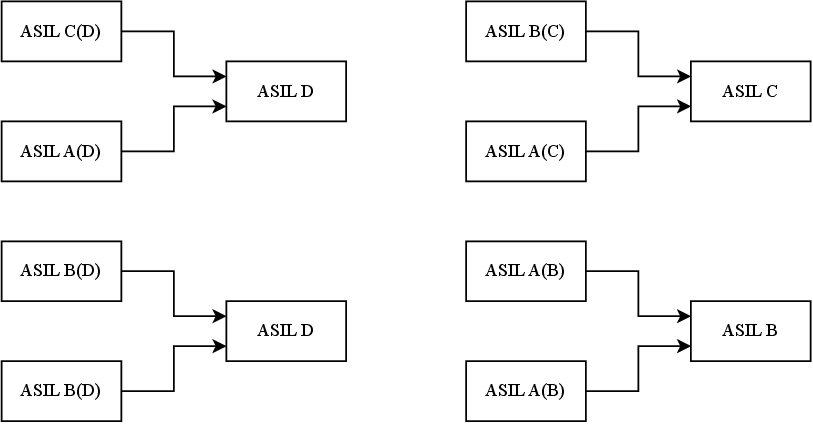}  %
  \caption{According to ISO 26262, it is possible to decompose a system architecturally into independent subsystems.}
  \label{fig:ASIL_decomposition}
\end{figure*}

The ISO~26262 standard sets out explicit requirements for functional safety in automotive systems, emphasising fault tolerance through architectural redundancy and ASIL decomposition. As shown in Figure \ref{fig:ASIL_decomposition}, different architectural patterns would be possible from an ASIL decomposition and redundancy viewpoint.

Conventional realizations of these principles rely on physically or logically independent redundant subsystems whose outputs are validated through runtime monitoring and voting mechanisms. In this work, we extend this paradigm to Transformer-based multimodal deep learning architectures by adapting these principles at the representational level using an encoder-decoder design.

Our proposed architecture consists of multiple modality-specific branches, each parameterized to extract high-level feature representations from heterogeneous sensor inputs (e.g., RGB images, LiDAR point clouds, monocular depth maps). Formally, let $\mathcal{X}_i$ denote the input space of modality $i$ and let $E_i: \mathcal{X}_i \rightarrow \mathcal{Z}$ be the corresponding branch mapping the raw input to a shared latent space $\mathcal{Z} \subseteq \mathbb{R}^d$. The latent embeddings are aggregated within $\mathcal{Z}$ via attention-based fusion mechanisms intrinsic to the Transformer architecture, yielding a unified, context-rich representation.

Downstream tasks are facilitated by modality-agnostic decoders $D_j: \mathcal{Z} \rightarrow \mathcal{Y}_j$, where $\mathcal{Y}_j$ denotes the target output space (e.g., semantic segmentation maps, 3D bounding boxes, driving commands). This explicit decoupling of modality-specific encoding from latent-space fusion and task-specific decoding enables modularity, controlled redundancy and verifiable independence between signal paths.

This approach achieves two principal safety-aligned benefits:
\begin{enumerate}
    \item \textbf{Intrinsic Redundancy:} If a sensor modality fails or becomes degraded, the remaining encoders continue supplying semantically meaningful features to the shared latent space. Provided that the remaining modalities contain sufficient task-relevant mutual information, the system maintains degraded but acceptable operational performance. This behavior is directly analogous to fail-operational redundant subsystems under ASIL decomposition.
    \item \textbf{Informational Enrichment:} When all modalities function nominally, their fusion in the latent space integrates complementary and partially orthogonal information streams. This redundancy at the information level improves the signal-to-noise ratio and mitigates both aleatoric and epistemic uncertainty, resulting in more robust perception and decision-making pipelines under diverse operational conditions.
\end{enumerate}

The independence of encoder branches structurally parallels the ISO~26262 requirement for freedom from common cause failures in redundant subsystems. Furthermore, the latent-space fusion mechanism provides an implicit runtime arbitration of multimodal signals, conceptually comparable to hardware-based voting and monitoring strategies found in traditional safety architectures.

In summary, the proposed multimodal Transformer architecture systematically embeds redundancy, diversity and fault tolerance at the representational level. By structurally mapping ISO~26262 safety principles to the design of state-of-the-art deep learning systems, this approach establishes a principled pathway for aligning large-scale AI models with established safety certification practices in the automotive domain.

We propose the following architecture:
\begin{figure*}[htb]
    \vspace{1em}
    \centering
    \includegraphics[width=0.96\textwidth]{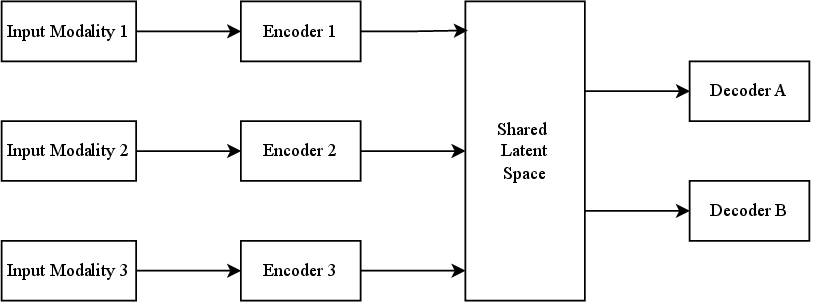}  %
  \caption{A multi-encoder-decoder architecture can be used to achieve multimodality in sensor data.}
  \label{fig:transformer_decomposition}
\end{figure*}


\section{Implementation}

In the proposed system, raw LiDAR point clouds are first projected onto the camera's image plane using calibrated extrinsic and intrinsic parameters. This generates a sparse depth map aligned with the camera’s viewpoint. Dedicated refinement stages then densify this sparse map to enhance the representation. The result is then spatially registered with the real camera feed to ensure consistent visual formatting across modalities.

A standard Vision Transformer tokenises this fused data, creating a unified input stream for the multimodal Transformer backbone.

Meanwhile, the Transformer-based language encoder independently processes textual data from driver requests, such as navigation commands or infotainment interactions. The visual and text tokens are then processed together by the downstream multimodal Transformer, which enables cross-modal reasoning.

This approach ensures that both sensing modalities contribute coherently to the shared latent space by converting LiDAR data into the same representational domain. This design enables robust fallback capabilities, ensuring consistent scene understanding even when camera input is degraded or unavailable. Crucially, this explicit mapping enables the approach to be seamlessly integrated with existing pretrained vision–language models without the need for architectural changes.


\section{Conclusion}

This work outlines a concept for integrating Transformer-based multimodal deep learning into automotive systems that is aligned with safety principles. The proposed architecture demonstrates how ISO 26262 principles can be systematically mapped to modern AI pipelines by explicitly designing for redundancy, sensor diversity and modularity. This paves the way for the safe deployment of large models in critical automotive applications.

While the conceptual framework presented here focuses on high-level architectural patterns, future work will address practical challenges such as verifying representational independence, developing monitoring strategies for latent-space fusion and devising scalable methods for validating fail-operational behaviour in real-world failure scenarios. Furthermore, the efficient deployment of multimodal Transformer architectures on embedded automotive platforms is still an area of ongoing research.

This research bridges the gap between cutting-edge deep learning and well-established functional safety practices, contributing towards the development of certifiable AI systems for autonomous driving and advanced driver-assistance systems (ADAS). We believe that continued convergence between safety engineering and AI research is essential to realise the full potential of intelligent vehicles safely and reliably.


\begin{credits}
\subsubsection{\ackname} This research was funded by the Federal Ministry of Research, Technology and Space of Germany (BMFTR) as part of the CeCaS project, FKZ: 16ME0800K
\end{credits}


\bibliographystyle{splncs04}
\bibliography{literature}

@book{ISO26262,
  title = {ISO 26262: Road vehicles – Functional safety},
  author = {{International Organization for Standardization}},
  year = {2018},
  publisher = {ISO},
  address = {Geneva, Switzerland}
}

@inproceedings{vaswani2017,
 author = {Vaswani, Ashish and Shazeer, Noam and Parmar, Niki and Uszkoreit, Jakob and Jones, Llion and Gomez, Aidan N and Kaiser, \L ukasz and Polosukhin, Illia},
 booktitle = {Advances in Neural Information Processing Systems},
 editor = {I. Guyon and U. Von Luxburg and S. Bengio and H. Wallach and R. Fergus and S. Vishwanathan and R. Garnett},
 pages = {},
 publisher = {Curran Associates, Inc.},
 title = {Attention is All you Need},
 url = {https://proceedings.neurips.cc/paper_files/paper/2017/file/3f5ee243547dee91fbd053c1c4a845aa-Paper.pdf},
 volume = {30},
 year = {2017}
}

@article{dosovitskiy2020,
  title={An Image is Worth 16x16 Words: Transformers for Image Recognition at Scale},
  author={Dosovitskiy, Alexey and Beyer, Lucas and Kolesnikov, Alexander and Weissenborn, Dirk and Zhai, Xiaohua and Unterthiner, Thomas and  Dehghani, Mostafa and Minderer, Matthias and Heigold, Georg and Gelly, Sylvain and Uszkoreit, Jakob and Houlsby, Neil},
  journal={ICLR},
  year={2021}
}

@INPROCEEDINGS{Chen2024,
  author={Chen, Long and Sinavski, Oleg and Hünermann, Jan and Karnsund, Alice and Willmott, Andrew James and Birch, Danny and Maund, Daniel and Shotton, Jamie},
  booktitle={2024 IEEE International Conference on Robotics and Automation (ICRA)}, 
  title={Driving with LLMs: Fusing Object-Level Vector Modality for Explainable Autonomous Driving}, 
  year={2024},
  volume={},
  number={},
  pages={14093-14100},
  doi={10.1109/ICRA57147.2024.10611018}}

@article{Yang2025, title={LiDAR-LLM: Exploring the Potential of Large Language Models for 3D LiDAR Understanding}, volume={39}, url={https://ojs.aaai.org/index.php/AAAI/article/view/33001}, DOI={10.1609/aaai.v39i9.33001}, abstractNote={Recently, Large Language Models (LLMs) and Multimodal Large Language Models (MLLMs) have shown promise in instruction following and image understanding. While these models are powerful, they have not yet been developed to comprehend the more challenging 3D geometric and physical scenes, especially when it comes to the sparse outdoor LiDAR data. In this paper, we introduce LiDAR-LLM, which takes raw LiDAR data as input and harnesses the remarkable reasoning capabilities of LLMs to gain a comprehensive understanding of outdoor 3D scenes. The central insight of our LiDAR-LLM is the reformulation of 3D outdoor scene cognition as a language modeling problem, encompassing tasks such as 3D captioning, 3D grounding, 3D question answering, etc. Specifically, due to the scarcity of 3D LiDAR-text pairing data, we introduce a three-stage training strategy and generate relevant datasets, progressively aligning the 3D modality with the language embedding of LLM. Furthermore, we design a Position-Aware Transformer (PAT) to connect the 3D encoder with the LLM, which effectively bridges the modality gap and enhances the LLM’s spatial orientation comprehension of visual features. Our experiments demonstrate that LiDAR-LLM effectively comprehends a wide range of instructions related to 3D scenes, achieving a 40.9 BLEU-1 score on the 3D captioning dataset, a Grounded Captioning accuracy of 63.1%, and a BEV mIoU of 14.3%.}, number={9}, journal={Proceedings of the AAAI Conference on Artificial Intelligence}, author={Yang, Senqiao and Liu, Jiaming and Zhang, Renrui and Pan, Mingjie and Guo, Ziyu and Li, Xiaoqi and Chen, Zehui and Gao, Peng and Li, Hongsheng and Guo, Yandong and Zhang, Shanghang}, year={2025}, month={Apr.}, pages={9247-9255} }

@inproceedings{Koopman2019,
author = {Koopman, Philip and Ferrell, Uma and Fratrik, Frank and Wagner, Michael},
title = {A Safety Standard Approach for Fully Autonomous Vehicles},
year = {2019},
isbn = {978-3-030-26249-5},
publisher = {Springer-Verlag},
address = {Berlin, Heidelberg},
url = {https://doi.org/10.1007/978-3-030-26250-1_26},
doi = {10.1007/978-3-030-26250-1_26},
booktitle = {Computer Safety, Reliability, and Security: SAFECOMP 2019 Workshops, ASSURE, DECSoS, SASSUR, STRIVE, and WAISE, Turku, Finland, September 10, 2019, Proceedings},
pages = {326–332},
numpages = {7},
location = {Turku, Finland}
}

@inproceedings{Frtunikj2014,
  title={Data-Centric Middleware support for ASIL assessment and decomposition in open automotive systems},
  author={Jelena Frtunikj and Michael Armbruster and Alois},
  year={2014},
  url={https://api.semanticscholar.org/CorpusID:17722274}
}

@INPROCEEDINGS{Singh2022,
  author={Singh, Amanpreet and Hu, Ronghang and Goswami, Vedanuj and Couairon, Guillaume and Galuba, Wojciech and Rohrbach, Marcus and Kiela, Douwe},
  booktitle={2022 IEEE/CVF Conference on Computer Vision and Pattern Recognition (CVPR)}, 
  title={FLAVA: A Foundational Language And Vision Alignment Model}, 
  year={2022},
  volume={},
  number={},
  pages={15617-15629},
  doi={10.1109/CVPR52688.2022.01519}}

@misc{li2019,
      title={VisualBERT: A Simple and Performant Baseline for Vision and Language}, 
      author={Liunian Harold Li and Mark Yatskar and Da Yin and Cho-Jui Hsieh and Kai-Wei Chang},
      year={2019},
      eprint={1908.03557},
      archivePrefix={arXiv},
      primaryClass={cs.CV},
      url={https://arxiv.org/abs/1908.03557}, 
}

@inproceedings{devlin2019,
    title = "{BERT}: Pre-training of Deep Bidirectional Transformers for Language Understanding",
    author = "Devlin, Jacob  and
      Chang, Ming-Wei  and
      Lee, Kenton  and
      Toutanova, Kristina",
    editor = "Burstein, Jill  and
      Doran, Christy  and
      Solorio, Thamar",
    booktitle = "Proceedings of the 2019 Conference of the North {A}merican Chapter of the Association for Computational Linguistics: Human Language Technologies, Volume 1 (Long and Short Papers)",
    month = jun,
    year = "2019",
    address = "Minneapolis, Minnesota",
    publisher = "Association for Computational Linguistics",
    url = "https://aclanthology.org/N19-1423/",
    doi = "10.18653/v1/N19-1423",
    pages = "4171--4186"
}

@book{pressman2009,
author = {Pressman, Roger},
title = {Software Engineering: A Practitioner's Approach},
year = {2009},
isbn = {0073375977},
publisher = {McGraw-Hill, Inc.},
address = {USA},
edition = {7}
}

@ARTICLE{LeCun1989,
  author={LeCun, Y. and Boser, B. and Denker, J. S. and Henderson, D. and Howard, R. E. and Hubbard, W. and Jackel, L. D.},
  journal={Neural Computation}, 
  title={Backpropagation Applied to Handwritten Zip Code Recognition}, 
  year={1989},
  volume={1},
  number={4},
  pages={541-551},
  doi={10.1162/neco.1989.1.4.541}}

\end{document}